# Seismic hazard assessment in Menton, France: Topographical site effect zoning considering a semi-empirical approach and a Machine Learning scheme

**Mathieu Allimant** – GERS/SRO, Univ. Gustave Eiffel, IFSTTAR, Marne la Vallée, France,
REPSODY Research team, CEREMA, Sophia Antipolis, France,
e-mail: mathieu.allimant@univ-eiffel.fr

**Etienne Bertrand –** GERS/SRO, Univ. Gustave Eiffel, IFSTTAR, Marne la Vallée, France,
e-mail : etienne.bertrand@univ-eiffel.fr

**Nathalie Glinsky** – REPSODY Research team, CEREMA, Sophia Antipolis, France,
e-mail: nathalie.glinsky@cerema.fr

**Celine Bourdeau** – GERS/SRO, Univ. Gustave Eiffel, IFSTTAR, Marne la Vallée, France,
e-mail : celine.bourdeau-lombardi@univ-eiffel.fr

**Abstract:** The presence of topography influences the seismic ground motion and may result in strong amplifications, generally at the top of hills and reliefs. The increasing urbanization of hills requires an accurate estimation of these effects even in areas of moderate seismicity. The simplified coefficients provided by the Eurocodes8 do not depend on the frequency and underestimate the amplification in many situations, which justifies the development of new methods based on easily accessible data. The city of Menton, located in the southeast of France, between the Alps and the Ligurian basin, is one of the most exposed metropolitan cities. We propose a study of topographic effects applied to the Menton area. Topographic amplification is calculated, on a wide frequency band, using the Frequency-Scaled Curvature method, from a DEM and an average value of the shear wave velocity. We then propose to apply an automatic clustering approach to classify the amplification curves into five groups with similar properties. We then deduce a first microzonation map of the topographic effects in the Menton area.

**Keywords:** Seismic ground motion, Site amplification, Topography, Microzonation with clustering

## 1. Introduction

The effects of topography on seismic ground motion have been observed during several earthquakes and have caused considerable damage. They consist of an amplification (sometimes very strong) or a de-amplification of the seismic signal associated with significant spatial variability of the surface motion. The increasing urbanization of hills requires an accurate estimation of these topographic effects even in a context of moderate seismicity. The Eurocodes8 addresses this issue in Annex A of Part 5 "Foundations, and Geotechnical Aspects" (Eurocodes8, 2005). It provides lump-sum coefficients for seismic action amplification used in slope stability study, which depend on a limited number of parameters (geometry, presence of a layer of soft medium). The maximum value being 1.4, it underestimates the amplification in many situations, as shown by De Martin (2012), and does not take into account the signal frequency dependence.
Despite numerous experimental and numerical studies conducted over the past 50 years (see for instance the review by Massa et al. (2014)), the parameters responsible for this complex phenomenon are still not completely identified. It is also difficult to separate in seismic recordings the contribution of the surface geometry from sub-soil properties and to select an amplification-free-reference station.

Parametric numerical studies have highlighted important parameters but very often underestimate the measured amplification levels. 3D simulations at a regional scale require the characteristics of the medium at depth, robust solvers and adapted computing resources. It is therefore of uttermost importance to turn to simplified techniques for estimating the amplification based on a limited number of data for microzonation studies. Such microzonation studies, proposing a division of a city into zones of similar seismic response, characterized by a specific response spectrum, are up to now mainly undertaken for lithological site effects.

In order to overcome this difficulty, a method was proposed by Maufroy et al. (2015) (Frequency-Scaled Curvature or "FSC" proxy) to estimate focusing and unfocusing effects of the waves due to the relief geometry. The authors propose a formula to estimate the amplification probability as a function of the curvature and the wavelength considered. This method requires little input data: a digital elevation model (DEM) and the average shear wave propagation velocity $V_s$ in the medium. The frequency aspect of the signal is taken into account through an appropriate smoothing of the curvature map. The FSC method has also been successfully applied to the Amatrice model (Mw 6.0, 2016) for comparison with the damage map (Maufroy et al., 2018). More recently, Wang et al. (2018) applied this same method to a numerical model of Hong Kong, to study the influence of a soil layer. We decided to apply and continue to explore this method in the context of a seismic risk estimation study of the site of Menton, one of the most seismically exposed cities in metropolitan France, because of a complex geodynamic context close to the Alps and the Ligurian basin. After a description of the site and its geology, we will apply the FSC method and present different maps and curves of the topographic amplification as a function of frequency. We will then propose preliminary results of classification of amplification curves intended to carry out a zonation of the topographical effects.

## 2. Topographical and geological context of the studied area

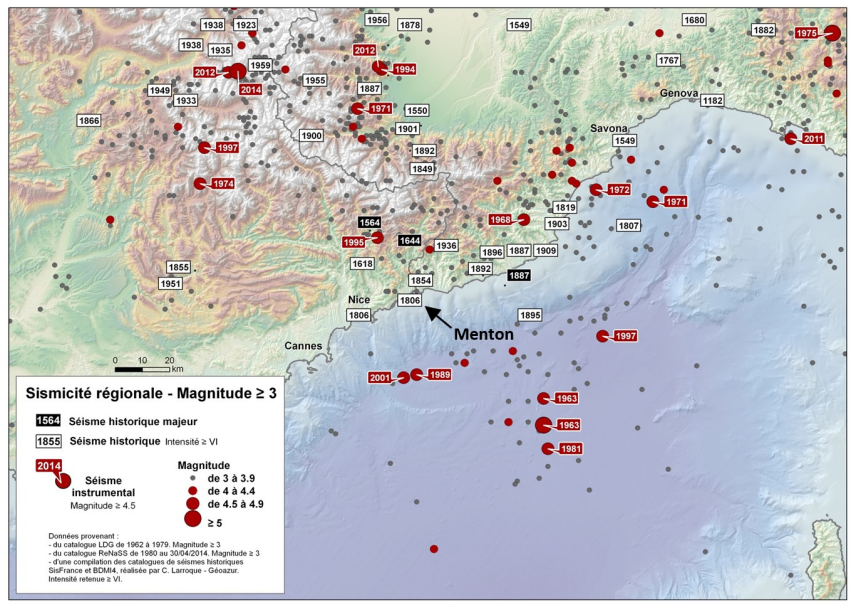

Fig. 1 - Seismicity in southeastern France (LDG, ReNass data for seismicity, adapted from GEOAZUR).

The city of Menton is located in the southeast of continental France, in the department of the Alpes Maritimes. It marks the end of the French Riviera at the border with Italy. In 2019 it counted almost 30 000 inhabitants (INSEE, French National Institute of Statistics and Economic Studies). According to the French seismic hazard zonation, dividing the territory into five hazard zones, Menton is located in a "moderate" seismic zone, characterized by a maximum acceleration on reference rock site of 0.16g to be considered for new constructions (class II of normal risk structures). The historical seismicity reveals 58 earthquakes recorded from 1920 up to 1999 (Larroque et al., 2001) with very few strong earthquakes Mw>6 and regular moderate earthquakes Mw<4 (Courboulex et al., 2003). The strongest recorded earthquakes are the 1887 Ligurian earthquake with Mw of 6.8-6.9 (Larroque et al., 2012) and the 1564 Roquebillière earthquake with an estimated MSK intensity of IX-X (Laurenti,1998). There is a permanent microseismicity (Fig. 1) with more than 5000 events recorded since 1980 by the BCSF (French Central Seismological Office), with epicenters all around Menton. The FSC method requires an estimate of the average shear wave velocity $V_s$ in the medium. Menton and its hinterland are characterized by Meso-Cenozoic sedimentary rocks as described in Fig. 2. At the scale of 1:50 000, 14 different lithologies can be observed. We propose to gather them in three main lithological groups, limestone, marl and sandstone, as explained in Fig. 3 (right). Values of the shear wave velocity are deduced from the compressional velocity $V_p$ in each group and their values are respectively set to 1700, 1200 and 800m/s (Table 1).

The morphology of the city consists of a series of hills interspersed with deep valleys that flow into the sea. The topography of the area is made up of steep reliefs that can reach an elevation of 1200m within 5km of the coastline, as shown in Fig. 3 (left). Strong topographical amplification is expected on this geomorphology. We use a Digital Elevation Model with a resolution of 5m from IGN (National Institute of Geographic and Forest Information).

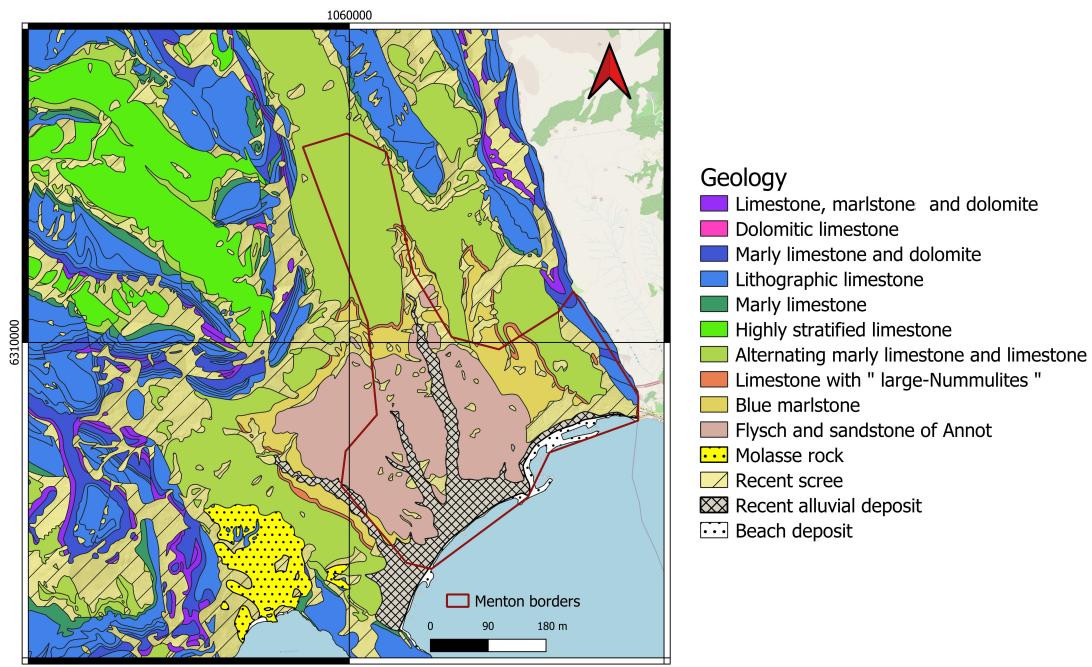

Fig. 2 - Geological map of the Menton area (RGF93 coordinates). The representation of the different lithological facies is based on the BRGM 1:50 000 data (infoterre.brgm.fr). The limits of the city of Menton are drawn in red.

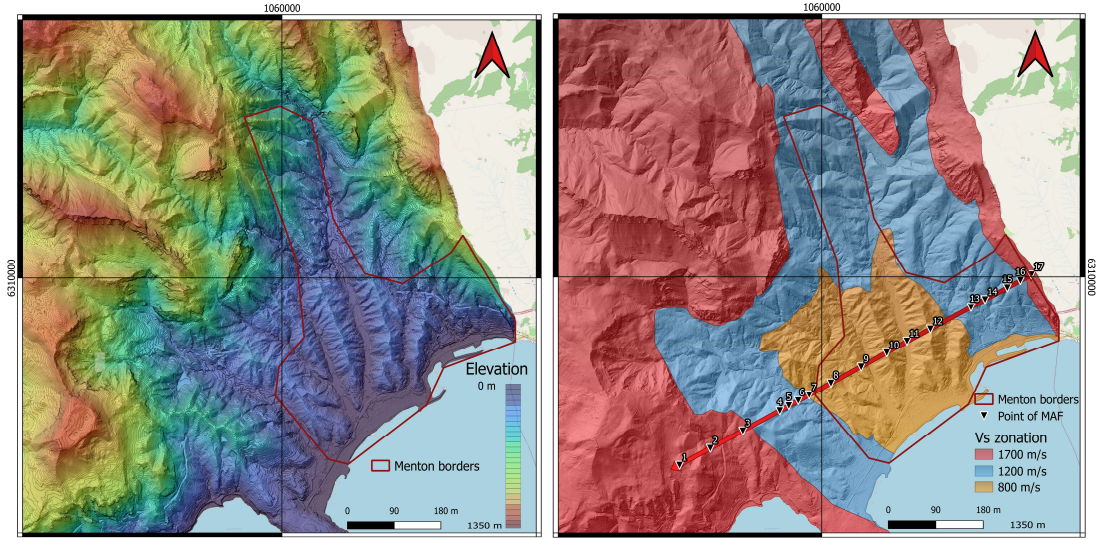

Fig. 3 - Left : View of DEM data at 5m spatial resolution. The color code ranges from blue for low altitudes to red for the highest reliefs. Right : Map of the $V_s$ zonation of the Menton region proposed for this study with the localization of the section and 17 points used in section 3.3.

Table 1. Approximation of $V_p$ and $V_s$ in the three main lithological groups. $V_s$ is deduced from $V_p$ by $V_p/V_s=\sqrt{3}$ (adapted from Astier, 1971).

| Lithology | $V_p$ (m/s) | $V_s$ (m/s) |
|---|---|---|
| Marls | 1800-2100 | 1050-1200 |
| Limestone | 3000-5000 | 1700-2800 |
| Sandstone | 800-3000 | 450-1700 |

## 3. Topographic amplification in the area of Menton

### 3.1. Description of FSC method

The Frequency Scaled Curvature method or "FSC method" (Maufroy et al., 2015) correlates the amplification due to the geometry of the topography (depending on the frequency) and the curvature of the relief. This method was developed with a database of numerical simulations of 200 earthquakes at Rustrel mountain (France). The authors considered a homogeneous medium, which allows them to focus on the amplifications due to the relief. The amplifications are computed from a dense network of 576 fictitious stations at the surface using the Median Reference Method "MRM" (Maufroy et al., 2012). The need for a "rock outcrop reference station" is avoided by taking the median or average motion at all the available stations as a reference. The analysis of the amplifications revealed a linear correlation between the curvature (i.e. the second derivative of the elevation map) and the topographic amplification. In practice, the matrix E containing the elevation values $E_{i,j}$ on a cartesian grid of interval h (both expressed in meters) is constructed from a DEM; i and j represent the indices along west-east and south-north directions. A second matrix C containing the curvature terms is deduced from E according to the method proposed by Zevenbergen & Thorne (1987); each component $C_{i,j}$ is calculated from the elevation value at (i,j) and at its four direct neighbors. For a given frequency f, the best correlation between topographic amplification and curvature was obtained when the curvature is smoothed along a characteristic length equal to the half-wavelength, $L_s = \lambda_s/2$ where $\lambda_s = V_s/f$. The smoothed curvature matrix $C_s$ is obtained by applying to C the smoothing operator consisting of a double convolution by a (n,n) matrix full of 1, centered around (i,j), and normalized by a factor $n^4$, n being necessarily an odd integer greater or equal to 3; the spatial extension of the smoothing window is n·h where n verifies $L_s=2·n·h$. Maufroy et al. (2015) then propose the following expression for the median value of the horizontal amplification factor (MAF) at each point (i,j) of the grid:

$$MAF(f) = 0.0008 \cdot \lambda s \cdot Cs(Ls) + 1 \qquad (1)$$

depending on the frequency, via the wavelength and the smoothed curvature. Note that values of MAF are close to one for small topographic curvature (i.e. C close to zero), whatever the frequency. The maximum frequency in the study is constrained by the resolution of the DEM. Since n is at least equal to 3 and $f=V_s/(4·n·h)$, it is possible to go up to 13Hz with $V_s$=800m/s and h=5m. We are interested in a set of frequencies between 0.5 and 10Hz with a step of 0.5Hz. The smoothing, depending on a value of n, integer and odd, the values of frequencies are not exactly those of the target sampling but those which approach it more for each value of $V_s$. The frequency sampling corresponding to each value of $V_s$ is given in Table 2; values are very close at low frequency; the maximum values correspond to n=3.

Two types of results are presented in the following: maps of amplification on the whole domain at a fixed frequency and amplification curves at selected points for all frequencies. At a fixed frequency, maps are realized by gathering on the same map MAF values corresponding to the $V_s$ value of the area. Note that, as explained above, the values of the frequency corresponding to each $V_s$ area are very close but not exactly the same (Table 2). We can reasonably make these maps since the differences between the frequency values in each area are very small.

*Table 2. Frequency sampling in relation to the different values of Vs.*

| $V_s$ 800m/s | $V_s$ 1200m/s | $V_s$ 1700m/s |
|---|---|---|
| 0.506 | 0.504 | 0.502 |
| 1.025 | 1.016 | 1.000 |
| 1.481 | 1.538 | 1.491 |
| 2.105 | 2.068 | 1.976 |
| 2.666 | 3.157 | 2.575 |
| 3.076 | 3.529 | 2.931 |
| 3.636 | 4.000 | 3.400 |
| 4.444 | 4.615 | 4.047 |
| 5.714 | 5.454 | 4.473 |
| 8.000 | 6.666 | 5.000 |
| 13.333 | 8.571 | 5.666 |
|  | 12.000 | 6.538 |
|  |  | 7.727 |
|  |  | 9.444 |
|  |  | 12.140 |

## 3.2. Maps of topographic amplification

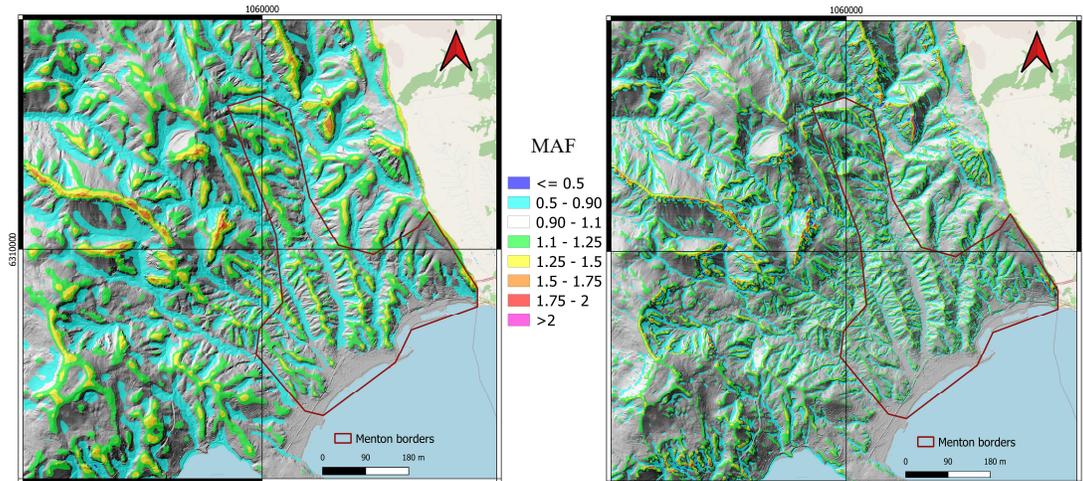

Fig. 4 - Maps of the topographic amplification (MAF) on the region of Menton obtained from a DEM at 5m resolution. The maps respect the color code from blue for de-amplification (MAF<1.0), grey for neutral areas (MAF ~1.0) and orange / red / pink for higher values of MAF.

We investigate here the spatial distribution of the topographical amplification on the whole domain. Fig. 4 shows an overlay of the DEM in 3D perspective, allowing to distinguish the hills from the valleys on which have been added the median amplification factors (MAF). Both figures correspond to frequencies respectively around 2.0Hz (left) and 8.0Hz (right). If we compare these two maps, morphologies with hills and ridges are all affected by topographic amplification factors ranging from 1.1 to more than 2.0 for the steepest ridges in the northwestern quarter of the map. In the southeastern quarter of the map which happens to be the downtown area of Menton, the hills of dwellings get a factor of amplification up to 1.4 at their crests. Conversely, the valleys reflect de-amplification effects (MAFs lower than 1.0). High values of amplification remain localized to the same reliefs comparing both plots, however the amplified surfaces are thinner at high frequency. The highest amplification values defined by the color code red / pink, are also more present on the ridges on the map corresponding to f=8.0Hz.

## 3.3. Topographic amplification curves at points of interest

In this section, we present amplification curves on the whole frequency band at points located on a profile that crosses downtown Menton, with points in different reliefs and different $V_s$ (Fig. 5). The profile is oriented west-east and shown in Fig. 3 (right). We observe very different levels of amplification, between 0.77, corresponding to de-amplification, and 1.33, a fairly high amplification, with a median at 1.04. There is a very high variability in the shapes of the curves. All the curves show strong variations as a function of frequency. We can classify the different curves into three main categories: curves with amplification on the whole frequency band and max(MAF)>1.2, points 10 and 12 (which are on hills), curves showing de-amplification for all frequencies and min(MAF)<0.9, points 2, 7, 11 and 13 (which are on valley, except point 2 which is on a complex relief) and mixed curves, alternating amplification and de-amplification according to the frequency, all other points (which are on concave or convex mid-slopes depending on the erosion of the relief). For this profile, the maximum amplification occurs between 2 and 4.0Hz.

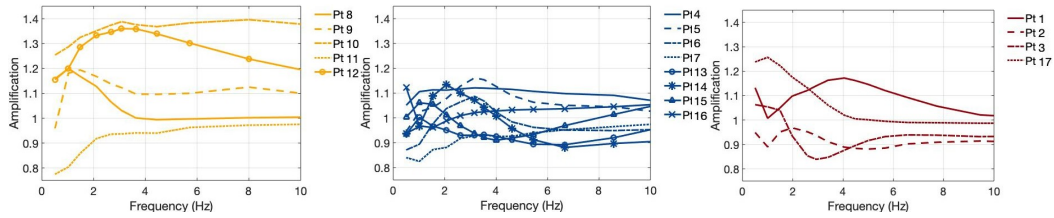

Fig. 5 - Evolution of the median amplification factor as a function of frequency along the cross section (Fig. 3, right). In orange for Vs 800m/s domain (left), blue for 1200m/s (center) and red for 1700m/s (right).

## 4. Microzonation of topographical site effect

### 4.1. K-Mean clustering

Clustering is one of the most common data mining techniques applied to extract knowledge from a large database (Ahmad & Dey, 2007). This technique is based on unsupervised learning algorithms, in which the objective is to divide a given set of objects into homogeneous groups, i.e clusters. Clustering has already been used in many applications (Bansal et al., 2017). Here, we use it to gather sites with a similar topographic seismic ground motion amplification curve. The general idea of clustering is to partition n objects of m-dimensions into k distinct clusters. In each cluster, the objects are more similar than the objects in the other clusters. In most common clustering algorithms, this similarity is evaluated using the Euclidean distance calculation. We use the K-Mean clustering algorithm included in Scikit-learn Python module (Pedregosa et al., 2011). Among the data, the K-Mean algorithm randomly selects first k objects as cluster centers. Then, each object in the database is assigned to the most similar center. The algorithm then computes the new mean for each cluster and reassigns each object to the nearest new center. This process iterates until no changes occur to the allocation of objects anymore (Bansal et al., 2017).

### 4.2. Two-steps approach

To investigate the feasibility of the approach, we focused on the city center of Menton, in the geological zone characterized with a $V_s$ of 800m/s. We used a two-step approach to first separate sites with amplification from sites without amplification or de-amplified. Then, we reprocessed the amplification curves alone to define subgroups of seismic amplification sites. To separate the sites that amplify the seismic signal from the others, a set of three clusters was considered (Fig. 6).

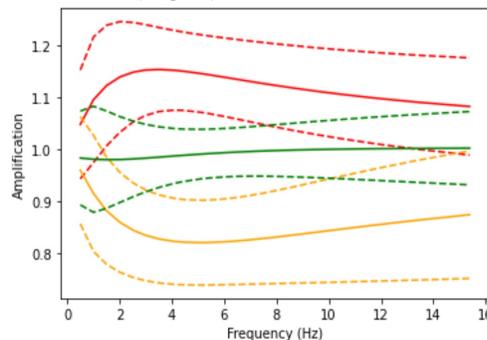

Fig. 6 - Result of the clustering in 3 groups. The red curves show a seismic amplification, the yellow ones characterize a de-amplification and the green ones correspond to sites where no significant seismic amplification is detected.

We see that the three groups are well separated but that the standard deviations are large (see dotted lines in Fig. 6). We note that the mean curve of the amplification cluster and the mean of the de-amplification group are quite symmetrical around the A=1 horizontal line. Here, the maximum of average amplification, equal to 1.15, is close to 3.5Hz.

To find the significant subgroups among the amplifying sites, a test on the number of clusters is first performed. This consists in calculating the total residual error for different numbers of clusters and choosing the smallest number of clusters offering sufficient accuracy.

Indeed, as the number of clusters increases, the total cost value (sum of squared distances to the closest centroid for all observations in the training set) decreases (Fig. 7, left). We finally consider clustering in 5 families as the sum of the distances is not significantly decreasing for a higher number of clusters. The result of this second computation, obtained after 25 iterations, is shown on Fig. 7 (right). The five cluster centers are notably different in shape and amplitude. The maximum amplification (1.3) is obtained for cluster 2 at around 4.0Hz.

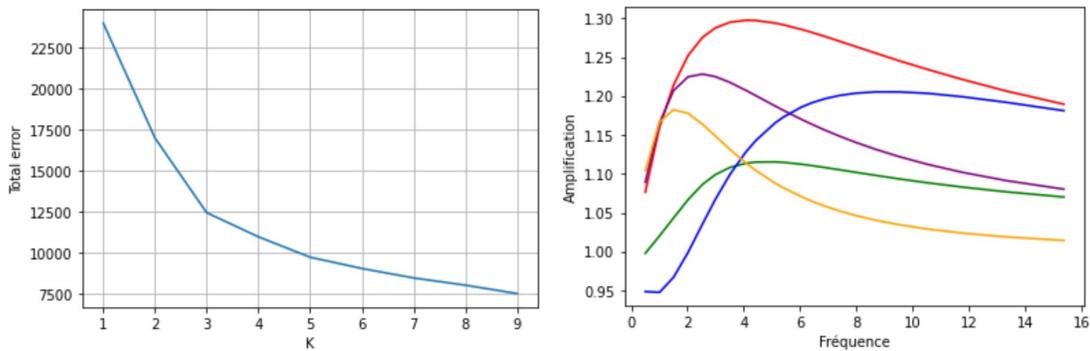

Fig. 7 - Left : Evolution of the total error with the number of considered clusters. Right : Result of the clustering of amplification curves in 5 groups.

All the average amplifications are lower than the maximum topographic amplification coefficient considered in the EC8 building code. However, the variability inside the clusters is still very large (Fig. 8). We notice that the amplification values inside a cluster, excepting cluster 1, is not normally distributed: the $84^{th}$ percentile does not match one standard deviation for all the investigated frequencies. Indeed, for cluster 2, these values are normally distributed only at frequencies lower than 3.0Hz.

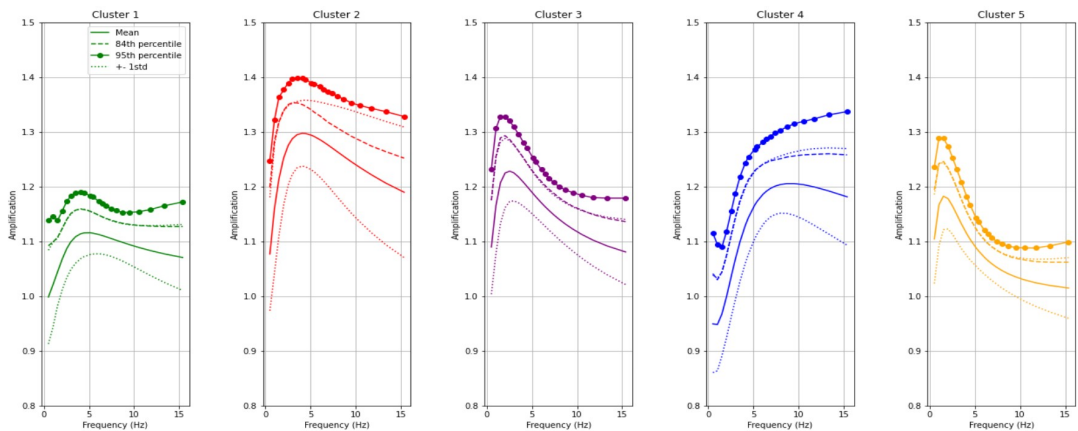

Fig. 8 - Seismic amplification curves for the 5 clusters. Average, ± 1 standard deviation, $84^{th}$ and $95^{th}$ percentiles.

For this cluster we observe some extreme amplifications above 4.0Hz for less than 5% of the sites, the maximum reaching almost 3.5 at 15.5Hz for one given site (Fig. 9). Keeping in mind that the individual amplification curve obtained with the FSC method is already an average (i.e. there is a 50% probability that sites will experience a higher amplification level) and derived from a smoother topographical context than in Menton, the seismic amplification finally associated with each cluster is still in question. For safety reasons we would suggest taking into consideration the 95$^{th}$ percentile as the adopted transfer function of the zone.

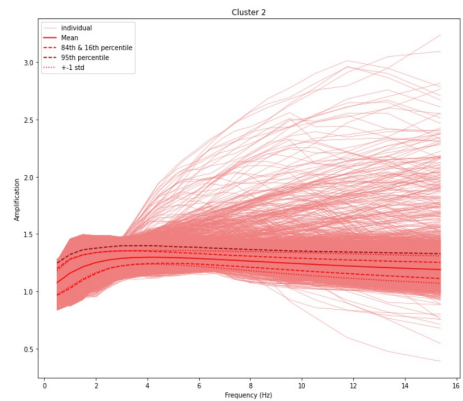

Fig. 9 - Distribution of seismic amplification curves for cluster 2.

### 4.3. Topographical site effect microzonation

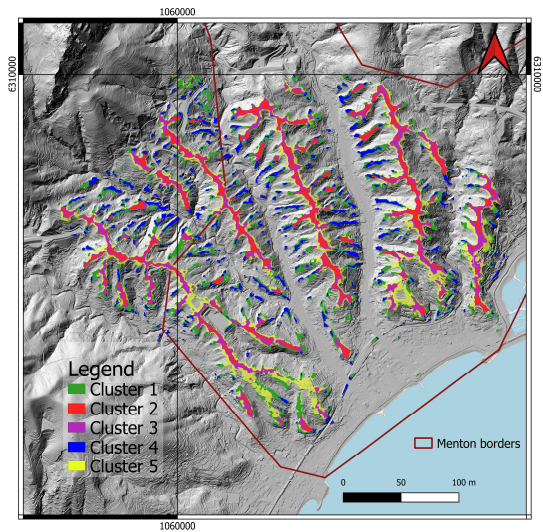

Fig. 10 - Microzonation of topographical site effects in the center of Menton.

Fig. 10 shows the geographical distribution of the 5 clusters of topographical seismic amplification in the center of Menton. As expected, most of the area is not affected by this kind of site effect. Furthermore, the cluster distribution follows a topographical logic: the areas with the highest amplification are located at the top of the hills and the most prominent hills show the highest amplification (from red to yellow). On the secondary reliefs (in green and blue), the amplification is less important and the peak amplification is at higher frequency (Fig. 8).

## 5. Conclusions

We estimated the seismic amplifications related to the topography in Menton (south-east of France) from the FSC method and proposed an innovative approach for microzonation cartography of the city center using an automatic clustering approach. Our preliminary results are very promising as the methodology provides a reliable assessment of the topographic site effect on a large area using only few and easily accessible input parameters. The microzonation procedure must still be extended to the whole studied area and the amplification functions need to be adapted to elastic response spectra. Furthermore, work in progress concerns the calibration of the FSC method on various topographies (smooth to steep) by 3D numerical simulations in various homogeneous and heterogeneous geologies.

## Acknowledgements

This work was carried out with the support of the French Ministry of Ecological Transition (MTE) within the framework of the Efftop2 project "Evaluation of the topographic site effect: numerical modeling on RAP-RESIF sites" (call for projects RAP-RESIF 2021). The authors would like to thank Emeline Maufroy and Agathe Roullé for fruitful discussions within the Efftop2 project about the FSC method and these results.

## References

-Ahmad, A., Dey, L. (2007) A k-mean clustering algorithm for mixed numeric and categorical data. Data & Knowledge Engineering, 63, 503–527.
-Astier, J-L. (1971). Géophysique appliquée à l'hydrogeologie. Masson & Cie.
-Bansal, A., Sharma, M., Goel, S. (2017) Improved K-mean Clustering Algorithm for Prediction Analysis using Classification Technique in Data Mining. JCA, 157, 35–40.
-Courboulex, F., Larroque, C., Deschamps, A., Gélis, C., Charreau, J., Stéphan, J.-F. (2003) An unknown active fault revealed by microseismicity in the south-east of France. Geophys. Res. Lett., 30.
-Eurocodes8 (2005) Design of structuresfor earthquake resistance – Part 5: Foundations, retaining structures and geotechnical aspects, NF-EN-1998-5, 41.
-Larroque, C., Béthoux, N., Calais, E., Courboulex, F., Deschamps, A., Déverchère, J., Stéphan, J.-F., et al. (2001) Active and recent deformation at the Southern Alps – Ligurian basin junction. Netherlands Journal of Geosciences, 80, 255–272.
-Larroque, C., Scotti, O., Ioualalen, M. (2012) Reappraisal of the 1887 Ligurian earthquake (western Mediterranean) from macroseismicity, active tectonics and tsunami modelling: Reappraisal of the 1887 Ligurian earthquake. Geophys. J. Int., 190, 87–104.
-Laurenti, A.. (1998). Les tremblements de terre des Alpes Maritimes. Serre.
-De Martin, F. (2012) Effets topographiques 2D en sismologie - Vérification des coefficients simplifiés Eurocode 8 par éléments spectraux. Tech. Rep. RP-61279-FR, BRGM, 47.
-Massa, M., Barani, S., Lovati, S. (2014) Overview of topographic effects based on experimental observations: meaning, causes and possible interpretations. Geophys. J. Int., 197, 1537–1550.
-Maufroy, E., Cruz-Atienza, V.M., Cotton, F., Gaffet, S. (2015) Frequency-Scaled Curvature as a Proxy for Topographic Site-Effect Amplification and Ground-Motion Variability. Bulletin of the Seismological Society of America, 105, 354–367.
-Maufroy, E., Cruz-Atienza, V.M., Gaffet, S. (2012) A Robust Method for Assessing 3-D Topographic Site Effects: A Case Study at the LSBB Underground Laboratory, France. Earthquake Spectra, 28, 1097–1115.
-Maufroy, E., Lacroix, P., Chaljub, E., Sira, C., Grelle, G., Bonito, L., Causse, M., Cruz-Atienza, V.,Hollender, F., Cotton, F., Bard, P.-Y (2018) Towards rapid prediction of topographic amplification at small scales : contribution of the FSC proxy and Pleiades terrain models for the 2016 Amatrice earthquake (Italy, Mw 6.0), 16th Conference on Earthquake Engineering, 13.
-Pedregosa, F., Varoquaux, G., Gramfort, A., Michel, V., Thirion, B., Grisel, O., Blondel, M., et al. (2011) Scikit-learn: Machine Learning in Python. JMLR 12, 2825-2830.
-Wang, G., Du, C., Huang, D., Jin, F., Koo, R.C.H., Kwan, J.S.H. (2018) Parametric models for ground motion amplification considering 3D topography and subsurface soils. Soil Dynamics and Earthquake Engineering, 115, 41–54.
-Zevenbergen, L.W., Thorne, C.R. (1987) Quantitative analysis of land surface topography. Earth Surf. Process. Landforms, 12, 47–56.